\documentclass[12pt]{article}
\usepackage{xcolor, soul}
\sethlcolor{yellow}
\usepackage{amsmath}
\usepackage{amssymb}
\begin{document}
\vskip 2cm
\begin{center}
{\bf {\Large Christ--Lee Model: (Anti-)Chiral Supervariable Approach to BRST Formalism}}\\

\vskip 3.2cm

{\sf B. Chauhan$^{(a)}$, $\,$ S. Kumar$^{(a)}$}\\\vskip 0.4 cm

{\it $^{(a)}$Physics Department,  Centre of Advance Studies, Institute of Science,}\\
{\it Banaras Hindu University, Varanasi - 221 005, (U.P.), India}\\

\vskip 0.3cm


\vskip 0.1cm

{\small {\sf {E-mails: bchauhan501@gmail.com; sunil.bhu93@gmail.com}}}

\end{center}

\vskip 3cm

\noindent
{\bf Abstract:} We derive  the  off-shell nilpotent of order two and absolutely anti-commuting
 Becchi-Rouet-Stora-Tyutin (BRST), anti-BRST and
(anti-)co-BRST symmetry transformations for the Christ--Lee (CL) model in one (0 + 1)-dimension (1D) 
of spacetime by exploiting the (anti-)chiral supervariable approach (ACSA) to  BRST formalism
where the {\it quantum} symmetry [i.e. (anti-)BRST along with (anti-)co-BRST] invariant quantities play a crucial role. 
We prove the nilpotency and absolute  anti-commutativity properties of the (anti-) BRST along with 
(anti-)co-BRST conserved charges within the scope  of ACSA to BRST formalism where we take
only {\it one} Grassmannian variable into account. We also show the (anti-)BRST and (anti-)co-BRST
invariances of the Lagrangian within the scope of ACSA. \\

\vskip 1.2cm



\newpage
\noindent
\section{Introduction}

\vskip 0.5 cm

Gauge theories describe  three (i.e. strong, weak, electromagnetic) out of four fundamental interactions of 
nature which are characterized by first-class constraints in the context  of Dirac's prescription 
for the classification scheme of constraints [1, 2]. The existence of the first-class constraints, in a given system, 
is the key signature  of a gauge theory. Many interesting  theories, in the domain of  physics, are expressed 
by the suitable Lagrangians that are invariant under the gauge symmetry transformations.  These symmetries 
are generated by the first-class constraints in a given gauge  theory.
For  the covariant canonical quantization of the gauge theory, the Becchi-Rouet-Stora-Tyutin
 (BRST) quantization procedure   plays a decisive  role where 
we replace the infinitesimal local gauge parameter by ghost and anti-ghost fields [3-6]. Thus, 
in this formalism, we have two fermionic  type global BRST  $(s_b)$ and anti-BRST $(s_{ab})$   transformations
at the {\it quantum} level (for a given local gauge symmetry  transformation at the classical level). 
These symmetry transformations are endowed with  two important  
 properties: (i) nilpotency   of order two (i.e. $s_{b} ^2 = 0,\; s_{ab}^2 = 0$), and  
(ii) absolute anti-commutativity (i.e. $s_b\,s_{ab} + s_{ab}\,s_{b} = 0$). 
The first property signifies  that theses quantum  BRST and  anti-BRST symmetry transformations 
are fermionic in nature whereas  second property shows that both symmetry transformations are linearly 
 independent of each other. Besides the (anti-)BRST symmetry transformations, we have two more fermionic and  linearly independent symmetry transformations which are christened  as the co-BRST ($s_d$) and anti-co-BRST ($s_{ad}$) symmetry transformations. The latter fermionic type  symmetry transformations are valid  for any D-dimensional $p$-form ($p = 1, 2, 3,...$) gauge  theories in $D = 2p$ dimensions of spacetime. We point out that there are some specific systems such as rigid rotor 
and Christ--Lee (CL) model in one-dimension (1D) that respect the  (anti-)co-BRST transformations along with 
to the (anti-)BRST transformations [7-10].

The geometrical interpretation and the emergence  of nilpotent (anti-)BRST symmetry transformations have been shown within
the ambit  of  Bonora-Tonin (BT) superfield formalism [11-13] where the Grassmannian variables $(\vartheta, \bar\vartheta)$
and their corresponding derivatives $(\partial_\vartheta, \partial_{\bar\vartheta})$ (with properties $\vartheta^2 = \bar\vartheta^2 = 0,
\;\partial_\vartheta ^2 = \partial_{\bar\vartheta}^2 = 0$ and  
$\vartheta\,\bar\vartheta + \bar\vartheta\,\vartheta = 0, \; \partial_{\vartheta}\,\partial_{\bar\vartheta} 
+ \partial_{\bar\vartheta}\,\partial_\vartheta = 0$)  play very important role. 
In BT-superfield approach, we see  the connections between the (anti-) BRST
symmetry transformations [$s_{(a)b}$] (with properties  $s_{b} ^2 = s_{ab}^2 = 0$ and $s_b\,s_{ab} + s_{ab}\,s_{b} = 0)$ and  Grassmannian translational generators $(\partial_\vartheta, \partial_{\bar\vartheta})$
 because of the fact that both have same algebraic structure. 
In this formalism, any D-dimensional 
Minkowskian manifold is generalized onto the (D, 2)-dimensional supermanifold. This suitably 
 {\it chosen} supermanifold  is denoted  
by the superspace coordinates $(x^\mu, \vartheta, \bar\vartheta)$ where $x^\mu (\mu  = 0, 1, 2,...,D-1)$ are the  spacetime coordinates
and $(\vartheta, \bar\vartheta)$ are a pair of Grassmannian variables.

The CL model is one of the simplest examples of gauge-invariant system
which is described by a singular Lagrangian [14].  Physically, the CL model represents   the motion of a  
point particle moving in a plane under the influence  of a central potential.
The CL model has been studied at the classical and quantum levels in different prospectives [14-20]. 
This model is endowed with  the first-class constraints 
in the Dirac's terminology for the classification scheme of constraints [1, 2]. 
This model is also quantized by using the Faddeev-Jackiw quantization 
where all the primary and derived constraints are treated on equal footing without any type of  further classification [20]. 
Within the framewok of BRST formalism, the CL model respects  {\it six} independent continuous 
 symmetries (i.e. BRST, anti-BRST, co-BRST, anti-co-BRST, ghost-scale and 
bosonic symmetries) (see, e.g. [9] for detail). The
BT-superfield formalism has been applied to obtain the absolutely anti-commuting and off-shell
 nilpotent (anti-)BRST as well as  (anti-)co-BRST
symmetry transformations where the technique of celebrated horizontality condition (HC) and dual
horizontality condition have been  used [10], respectively.

In our recent set of  papers [21-25], we have used a {\it newly} proposed formalism which has been called by us   as 
 the (anti-)chiral superfield/supervariable approach (ACSA) to BRST 
formalism. In this approach, we take into account only one  Grassmannian variable  
in the expression for the superfield/supervariable. Thus, the resulting superfield/supervariable turns into (anti-)chiral version
 of the superfield/supervariable. In other words, in this formalism, any D-dimensional Minkowskian
 manifold is generalized onto  (D, 1)-dimensional supersub-manifolds  of the  most {\it general}  (D, 2)-dimensional supermanifold.
The proof of the absolute anti-commutativity property of Noether's conserved charges is obvious 
in the case of BT-superfield formalism where the full super expansions of superfields/supervariables  are taken into account.  
In the case of ACSA, we have also been able to show the nilpotency and absolute anti-commutativity properties
of conserved charges despite the fact that we have taken only one Grassmannian variable into account. 
In our present endeavor, we derive the (anti-)BRST together with  (anti-)co-BRST symmetry 
transformations where some specific sets of (anti-)BRST and (anti-)co-BRST invariant restrictions
play very important role. We also show the   absolute anti-commutativity as well as  nilpotency  
properties of (anti-)BRST and (anti-) co-BRST conserved charges within the realm of 
ACSA to BRST formalism.

Against the background  of the above paragraph, it has been found that the 
nilpotency of the ${\cal N} = 2$ super charges are true for any 
${\cal N} = 2$ supersymmetric (SUSY) quantum mechanical models (see, e.g. [26-28]) within the ambit of 
(anti-)chiral supervariable approach  to BRST formalism. 
However, the application of ACSA, in the realm of ${\cal N} = 2$ SUSY quantum mechanical model, does not lead 
to the absolute anti-commutativity of the ${\cal N} = 2$ super conserved charges. Rather, it has been found that  
the anti-commutator of  the above ${\cal N} = 2$ SUSY  conserved charges leads to  
the time translation of the variable on which it acts.  
Thus, it is crystal clear that, ACSA to BRST formalism does not lead to the derivation of 
absolute anti-commutativity of the charges for all ${\cal N} = 2$ SUSY theories.

The different sections of our present paper are arranged  as follows. In Sec. 2, we discuss
the (anti-)BRST  and (anti-)co-BRST symmetry transformations  for the CL model
and derive the conserved charges. Our Sec. 3 deals with the ACSA to BRST formalism where
 we derive the (anti-)BRST  symmetry transformations. Sec. 4  is devoted
 to the derivation of (anti-)co-BRST  symmetry transformations by using the ACSA to BRST formalism where the super expansions of 
(anti-)chiral supervariables  are utilized in a fruitful manner. In Sec. 5, we express the conserved (anti-)BRST  and (anti-)co-BRST  charges on
the (1, 1)-dimensional supersub-manifolds [of the most {\it general}  (1, 2)-dimensional supermanifold]
 on which our theory is generalized and provide the proof of nilpotency and absolute
anti-commutativity properties of the (anti-)BRST along with  (anti-)co-BRST  charges within the ambit  of ACSA
to BRST formalism. In Sec. 6, we discuss the (anti-)BRST and (anti-)co-BRST  invariances of the Lagrangian
within the scope  of ACSA. Finally, we point out our
key results and discovery  in Sec. 7 and mention a few future scopes  for further investigation.\\

\noindent
\section{Preliminaries: Symmetries and their Corresponding Generator for the  Christ--Lee Model}

The first-order and gauge-invariant Lagrangian of the 
 Christ--Lee (CL) model\footnote{The other equivalent second-order Lagrangian ($L_s$) [14] associated with the 
Christ--Lee model is $L_s = \frac{1}{2}\,\dot r^2  + \frac{1}{2}\,\dot r^2 \, (\dot \varphi   - z)^2 - V(r)$. 
But we choose only the first-order Lagrangian $L_f$ in our present work because it respects maximum  
symmetry transformations.} in  $(0 + 1)$-dimension (1D) of spacetime in polar 
coordinates system is given by [14, 16, 19], 
\begin{eqnarray}
&& L_f = \dot r \,p_r + \dot \varphi \, p_\varphi  - \frac{1}{2}\, p^2_r - \frac{1}{2 r^2}\, p^2_\varphi  - z \,p_\varphi  - V(r),
\end{eqnarray}
where $\dot r$ and  $\dot\varphi $ are the generalized velocities, $p_r$ and  $p_\varphi $ are their  corresponding canonical 
momenta, respectively and $z$ is a Lagrange multiplier which enforces a constraint $p_{\varphi } \approx 0$. 
This Lagrangian explains that a two dimensional particle 
moves under  the influence of  the  central potential $V(r)$ bounded from below.

The above system has a primary constraint as follow
\begin{eqnarray}
\Phi_1 = \frac{\partial L_f}{\partial {\dot z}} = p_z \approx 0.
\end{eqnarray}
The time derivative  of the primary constraint $\Phi_1$ leads to the following secondary  constraint
\begin{eqnarray}
\frac{d \Phi_1}{d \tau}  = \frac{d}{d \tau} \Big(\frac{\partial L_f}{\partial {\dot z}}\Big) \approx 0 \Rightarrow  \Phi_2 = p_\varphi  \approx 0.
\end{eqnarray}
It is clear that both $\Phi_1$ and $\Phi_2$ are first-class constraints.  
The gauge symmetry transformation generator can be written in terms of first-class constraints  as  
\begin{eqnarray}
G = {\dot \chi} (\tau)\,\Phi_1 + \chi (\tau)\, \Phi_2,
\end{eqnarray}
where $\chi(\tau)$ is an infinitesimal and time dependent  local gauge parameter and 
${\dot \chi} (\tau) = {d\chi}/{d\tau}$. Using the definition of a generator
\begin{eqnarray}
\delta \;\phi (\tau) = - i\, [ \phi (\tau),\; G], \qquad \phi  = r, p_r, \varphi , p_\varphi , z,   
\end{eqnarray}
where $\phi$ is the generic variable that is  present in the first-order  Lagrangian $L_f$.
We deduce  the following  local  gauge  transformations by exploiting  Eq. (5), namely; 
\begin{eqnarray}
\delta\, z (\tau) = \dot \chi (\tau), \qquad  \delta \,\varphi  (\tau) = \chi (\tau), \qquad  \delta[r (\tau),\, p_r (\tau),\, p_\varphi  (\tau), \; V(r)] = 0.
\end{eqnarray}
It is elementary  to check that, under the above local   gauge symmetry transformations, the Lagrangian under consideration  
remains invariant (i.e. $\delta L_f = 0 $). 


The (anti-)BRST invariant Lagrangian for the (0 + 1)-dimensional  CL model containing the gauge-fixing   
and Faddeev-Popov ghost terms  is  given  by [15]
\begin{eqnarray}
L = \dot r \,p_r + \dot \varphi \, p_\varphi  - \frac{1}{2}\, p^2_r - \frac{1}{2 r^2}\, p^2_\varphi  - z \,p_\varphi  - V(r) 
+  \frac{1}{2}\,{\cal B} ^2 +  {\cal B} \,(\dot z + \varphi ) + i \, \bar C \, C  - i\,\dot {\bar C}\, \dot C,
\end{eqnarray}
where the Nakanishi--Lautrup type {\it auxiliary} variable ${\cal B}$ is used to linearize the gauge-fixing term and 
 the Faddeev--Popov (anti-)ghost variables $(\bar C)C$  are used to make the Lagrangian BRST invariant.
These fermionic variables $(\bar C)C$ (with $C^2 = \bar C^2 = 0, C\bar C + \bar C C = 0$) have ghost numbers $(-1)+1$, respectively.
The above Lagrangian respects the following  off-shell nilpotent [i.e. $s_{(a)b} ^2 = s_{(a)d}^2 = 0$]
and    absolutely anti-commuting (i.e. $s_b\, s_{ab} + s_{ab}\, s_{b} = 0$ and $s_d\, s_{ad} + s_{ad}\, s_d = 0$)  (anti-)BRST 
along with (anti-)co-BRST   symmetry transformations:  
\begin{eqnarray}
&& s_{ab} z = \dot {\bar C}, \quad\; s_{ab} \varphi  = \bar C, \quad \;s_{ab}\, C = - \,i \,{\cal B},  
\quad s_{ab}[r, \,p_r,\, p_\varphi , {\cal B},  \, \bar C] = 0,\nonumber\\
&& s_{b} z = \dot C, \;\;\quad s_{b} \varphi  = C, \;\quad\; s_{b} \bar C = i\,{\cal B},  \qquad \;\; s_b[r, \,p_r,\, p_\varphi ,\, {\cal B},  \,C] = 0, 
\end{eqnarray}
\begin{eqnarray}
&& ~~~~s_{ad}\, z = C, \quad s_{ad}\, \varphi  = - \dot{C}, \quad s_{ad} \, \bar C = 
- i\, p_\varphi , \quad s_{ad}\,[r,\, p_r,\, p_\varphi ,\, {\cal B}, \, C] = 0,\nonumber\\
&& ~~~~s_d\, z = \bar C, \;\quad s_d\, \varphi  = - \dot{\bar C}, \;\;\quad s_d \,C = 
 i\, p_\varphi , \qquad s_d\,[r,\, p_r,\, p_\varphi ,\, {\cal B}, \, \bar C] = 0.
\end{eqnarray}
It can be  clearly checked that under the above (anti-)BRST [Eq. (8)] and (anti-)co-BRST [Eq. (9)]
 symmetry transformations  the Lagrangian [Eq. (7)] remains quasi-invariant (i.e. modulo a total time derivative):
 \begin{eqnarray}
&& s_{b} \,L = \frac{d}{d \tau}\big({\cal B}\, \dot C\big), \;\,\qquad\qquad s_{ab} \,L = \frac{d}{d \tau}\big({\cal B}\, \dot {\bar C}\big),\nonumber\\
&& s_d \,L = - \frac{d}{d \tau}\big(p_\varphi \, \dot {\bar C}\big), \quad\qquad s_{ad} \,L 
= - \frac{d}{d \tau}\big(p_\varphi \, \dot C\big).   
\end{eqnarray}
As a result, the action integral $S = \int d \tau\, L$ remains invariant under the (anti-)BRST as well as  
(anti-)co-BRST symmetry transformations 
[i.e. $s_{(a)b}\,S = 0, \; s_{(a)d}\,S = 0$]. According to Noether's theorem, the invariance of 
the above Lagrangian under the  nilpotent (anti-)BRST together with (anti-)co-BRST symmetry  
transformations leads to the following  (anti-)BRST charges  $[{\cal Q}_{(a)b}]$ and 
(anti-)co-BRST charges $[{\cal Q}_{(a)d}]$, namely; 
\begin{eqnarray}
&&{\cal Q}_{ab}  = \; {\cal B} \, \dot {\bar C} + p_\varphi \, \bar C \, \; \equiv\,\;  {\cal B} \, \dot {\bar C} - \dot {\cal B} \, {\bar C}, \nonumber\\
&&{\cal Q}_{b} \;\, = \; {\cal B} \, \dot C + p_\varphi  \, C \,\; \equiv\,\;  {\cal B} \, \dot C - \dot {\cal B} \, C,
\end{eqnarray}
\begin{eqnarray}
 &&{\cal Q}_{ad} = \;  {\cal B} \,  C - p_\varphi \, \dot C \,\; \equiv\,\;   {\cal B} \, C + \dot {\cal B} \, \dot C, \nonumber\\
&&{\cal Q}_d \;\, = \; {\cal B} \, \bar  C - p_\varphi  \, \dot {\bar C} \,\; \equiv\,\;  {\cal B} \, \bar C + \dot {\cal B} \, \dot {\bar C},
\end{eqnarray}
where the equivalent forms of the above charges are written
 with the help of the equation of motion: $p_\varphi  = -\,\dot {\cal B}$.
The above  charges are nilpotent of order two  [i.e. ${\cal Q}_{(a)b}^2 = {\cal Q}_{(a)d}^2 = 0$] 
and  anti-commuting in nature (i.e. ${\cal Q}_b\, {\cal Q}_{ab} + {\cal Q}_{ab}\,{\cal Q}_{b} = 0$ 
and ${\cal Q}_d\, {\cal Q}_{ad} + {\cal Q}_{ad}\,{\cal Q}_d = 0$). The conservation law for these
 charges [i.e. $\frac {d}{d\tau} {\cal Q}_{(a)b} = 0$ 
and $\frac {d}{d\tau} {\cal Q}_{(a)d} = 0$] can be easily  proven by
using the following interesting Euler-Lagrange equations of motion\footnote{Besides these EOMs, we use 
a equation $\ddot {\cal B}  + {\cal B} = 0$ derived from the EOMs (13) to prove the conservation law for 
 the (anti-)BRST together with  (anti-)co-BRST conserved  charges [Eqs. (11), (12)].} (EOMs) derived from 
Lagrangian $L$ of our theory  [Eq. (7)], namely;   
\begin{eqnarray}
&&\dot {\cal B} + p_\varphi  = 0, \quad {\cal B} = \dot p_\varphi , \quad {\cal B} = - (\dot z + \varphi ),  \quad \dot p_r - \frac{p^2_\varphi }{r^3} + V'(r) = 0, \nonumber\\
&& \dot r = p_r, \quad \dot \varphi  - z - \frac{p_\varphi }{r^2} = 0, \quad  \bar C +  \ddot {\bar C} = 0, \qquad  C + \ddot C = 0.
\end{eqnarray}
The (anti-)co-BRST  and (anti-)BRST conserved charges are the generators of the 
(anti-) co-BRST and (anti-)BRST symmetry transformations, respectively. As one can easily check  that
following relationships are true 
\begin{eqnarray}
&&s_d \,\xi    = - i \,\big[\xi  ,\; {\cal Q}_d \big]_\pm,  \qquad s_{ad} \,\xi   = - i \,\big[\xi   , \;{\cal Q}_{ad}\big]_\pm, \nonumber\\
&&s_{b} \,\xi    = - i \,\big[\xi   ,\; {\cal Q}_{b} \big]_\pm,  \qquad s_{ab} \,\xi   = - i \,\big[\xi   , \;{\cal Q}_{ab}\big]_\pm,
\end{eqnarray}
where $\xi   $ denotes  any generic variable present in the Lagrangian $L$ of our theory. The  
subscript $(\pm)$  on the square brackets denotes the (anti)commutator which depend  on the
nature of  generic variables $\xi $ being (fermionic)bosonic in nature.\\

\section{Nilpotent Quantum  (Anti-)BRST Symmetry Transformations: (Anti-)Chiral Supervariable Approach}

In this section, we determine  the  nilpotent (anti-)BRST symmetry transformations [cf. Eq. (8)]  by using  (anti-)chiral 
supervariable approach (ACSA) to BRST
formalism where we shall use the (anti-)chiral super  expansions of supervariables. 
Towards this goal, first of all, we generalize the ordinary variables  of the Lagrangian (7) onto (1, 1)-dimensional
{\it anti-chiral} supersub-manifold [of the most {\it common} (1, 2)-dimensional supermanifold] as follows, 
\begin{eqnarray}
&&z (\tau) \, \longrightarrow \;{\cal Z} (\tau, \bar\vartheta)  \;= \; z (\tau) + \bar\vartheta\,f_1 (\tau),\nonumber\\
&& \varphi  (\tau)\,  \longrightarrow \;\Theta (\tau, \bar\vartheta) \; = \; \varphi  (\tau) + \bar\vartheta\,f_2 (\tau), \nonumber\\
&&C(\tau) \longrightarrow \; F (\tau, \bar\vartheta)  \;=\;  C(\tau) + i\,\bar\vartheta\,b_1 (\tau),\nonumber\\
&& \bar C (\tau) \longrightarrow \; \bar F (\tau, \bar\vartheta) \; = \;  \bar C (\tau) + i\,\bar\vartheta\,b_2 (\tau), \nonumber\\
&& r (\tau) \;\longrightarrow \; R (\tau, \bar\vartheta)  \; = \;  r (\tau) + \bar\vartheta\,f_3 (\tau), \nonumber\\
&& p_r (\tau) \longrightarrow \; P_r (\tau, \bar\vartheta)  = \; p_r (\tau) + \bar\vartheta\,f_4 (\tau), \nonumber\\
&& p_\varphi  (\tau) \longrightarrow \; P_\varphi  (\tau, \bar\vartheta)  = \;  p_\varphi  (\tau) + \bar\vartheta\,f_5 (\tau), \nonumber\\ 
&&  {\cal B} (\tau) \; \,\longrightarrow \; {\tilde  {\cal B}}(\tau, \bar\vartheta) \, = \;   {\cal B} (\tau) + \bar\vartheta\,f_6 (\tau), 
\end{eqnarray}
where $b_1, b_2$ are the bosonic  derived variables  and $f_1, f_2, f_3, f_4, f_5, f_6$ are the fermionic
derived variables due to  fermionic nature of $\bar\vartheta$. We determine the precise  value of these
derived variables in terms of the {\it auxiliary} and  {\it basic}  variables 
present in the BRST invariant Lagrangian (7) by using the BRST invariant quantities/restrictions.

According to the basic principles  of ACSA, the BRST invariant
quantities  {\it must} remain independent of the Grassmannian variable ($\bar\vartheta$) when they 
are generalized onto the (1, 1)-dimensional   {\it anti-chiral}   supersub-manifold. The BRST invariant quantities are the
specific combinations of the variables present in Lagrangian (7). These are given as follows
\begin{eqnarray}
&& s_{b} (r, p_r, p_\varphi , {\cal B},  C) = 0, \quad s_{b} (z\, \dot C) = 0,\quad s_{b} (\varphi \, C) = 0,\nonumber\\
&& s_{b} (\dot {\cal B} \, z + i\, \dot {\bar C}\,\dot C) = 0, \quad s_{b} (\dot\varphi  - z) = 0,\quad s_{b} ({\cal B}\,\varphi  + i\,\bar C\, C) = 0.
\end{eqnarray}
We generalize the above BRST invariant restrictions  onto the (1, 1)-dimensional 
{\it anti-chiral} supersub-manifolds (of the suitably chosen most {\it common} (1, 2)-dimensional supermanifold) 
\begin{eqnarray} 
&& R (\tau, \bar\vartheta)  = r(\tau), \; P_r (\tau, \bar\vartheta)  = p_r (\tau), \;  P_\varphi  (\tau, \bar\vartheta)  = p_\varphi  (\tau),\;
 {\tilde {\cal B}} (\tau, \bar\vartheta)  = {\cal B} (\tau), \;  F (\tau, \bar\vartheta)  = C(\tau), \nonumber\\
 &&{\cal Z} (\tau, \bar\vartheta) \,\dot F (\tau, \bar\vartheta)  = z (\tau)\, \dot C(\tau), \quad \Theta (\tau, \bar\vartheta) \, F (\tau, \bar\vartheta)  
= \varphi  (\tau)\, C(\tau), \nonumber\\
&&{\dot {\tilde {\cal B}}} (\tau, \bar\vartheta)  \,{\cal Z} (\tau, \bar\vartheta)  +  i\,\dot{\bar F} (\tau, \bar\vartheta) \, \dot C (\tau, \bar\vartheta) 
= \dot  {\cal B} (\tau) \, z (\tau) + i\, \dot {\bar C} (\tau)\,\dot C (\tau), \nonumber\\
&& \dot\Theta  (\tau, \bar\vartheta)  - {\cal Z} (\tau, \bar\vartheta)  = \dot\varphi  (\tau) - z (\tau),\nonumber\\ 
&&{\tilde  {\cal B}} (\tau, \bar\vartheta)  \,\Theta (\tau, \bar\vartheta)  + i\,{\bar F} (\tau, \bar\vartheta) \,  F (\tau, \bar\vartheta) 
=  {\cal B} (\tau) \, \varphi  (\tau) + i\,  {\bar C} (\tau)\, C (\tau). 
\end{eqnarray} 
The above restrictions lead to the derivation of  the derived variables
in terms of the {\it basic} and {\it auxiliary} variables. To determine the value of 
these  variables, we perform the step-by-step explicit calculations. For this purpose,  
first of all, we use the generalization of the {\it trivial} BRST invariant restrictions
given in the first line of Eq. (17) as: 
\begin{eqnarray} 
&& P_\varphi  (\tau, \bar\vartheta)  = p_\varphi  (\tau) \Longrightarrow f_5 = 0, 
\qquad \; \, {\tilde  {\cal B}} (\tau, \bar\vartheta)  
= {\cal B} (\tau) \Longrightarrow f_6 = 0, \nonumber\\
&& R (\tau, \bar\vartheta) \; = r(\tau) \; \;\Longrightarrow f_3 = 0, \qquad\;\, F (\tau, \bar\vartheta) 
 = C(\tau) \Longrightarrow b_1 = 0,\nonumber\\
&& P_r (\tau, \bar\vartheta)  = p_r (\tau) \; \Longrightarrow f_4 = 0.
\end{eqnarray} 
After substituting the above value of derived variables from (18) to (15), we get the following expressions 
for the   {\it anti-chiral}   supervariables, namely;
\begin{eqnarray}
&&C(\tau) \longrightarrow \; F ^{(b)} (\tau, \bar\vartheta)  = C(\tau) + \bar\vartheta\,(0) \; \equiv C(\tau) + \bar\vartheta\,[s_b\, C (\tau)] ,\nonumber\\
&& r (\tau) \; \longrightarrow \; R ^{(b)} (\tau, \bar\vartheta)  = r (\tau) + \bar\vartheta\,(0) \;\; \equiv r (\tau) + \bar\vartheta\,[s_b\, r (\tau)], \nonumber\\
&& p_r (\tau) \longrightarrow \; P_r  ^{(b)} (\tau, \bar\vartheta)  = p_r (\tau) + \bar\vartheta\, (0) \equiv p_r (\tau)
 + \bar\vartheta\, [s_b\, p_r (\tau)], \nonumber\\
&& p_\varphi  (\tau) \longrightarrow \; P_\varphi  ^{(b)} (\tau, \bar\vartheta)  =  p_\varphi  (\tau) + \bar\vartheta\, (0)
 \equiv p_\varphi  (\tau) + \bar\vartheta\, [s_b\, p_\varphi  (\tau)], \nonumber\\ 
&&  {\cal B} (\tau) \; \longrightarrow \; {\tilde  {\cal B}} ^{(b)} (\tau, \bar\vartheta)  =   {\cal B} (\tau) + \bar\vartheta\,(0)\; \equiv  {\cal B} (\tau) + \bar\vartheta\,[s_b\, {\cal B} (\tau)], 
\end{eqnarray}
where the superscript $(b)$ on the   {\it anti-chiral}   supervariables  denotes that these supervariables  have been obtained
 after the use of BRST invariant quantities. It is clear that the coefficients of Grassmannian variable $\bar\vartheta$ are simply  the quantum BRST symmetries (8). Now, in the case of {\it non-trivial} BRST invariant restrictions: $s_{b} (z\, \dot C) = 0$ and $s_{b} (\varphi \, C) = 0$,  the following generalizations  onto (1, 1)-dimensional supersub-manifold, namely;  
\begin{eqnarray} 
 &&{\cal Z} (\tau, \bar\vartheta) \,\dot F ^{(b)} (\tau, \bar\vartheta)  = z (\tau)\, \dot C(\tau), 
\qquad \Theta (\tau, \bar\vartheta) \, F ^{(b)} (\tau, \bar\vartheta)   = \varphi  (\tau)\, C(\tau), 
\end{eqnarray} 
lead to the following interesting  results 
\begin{eqnarray} 
f_1 (\tau)\,\dot C(\tau) = 0 \quad & \Longrightarrow & \quad f_1 (\tau) \propto \dot C(\tau), \nonumber\\
& \Longrightarrow & \quad f_1 (\tau)  =  \kappa _1 \,\dot C(\tau), \nonumber\\
f_2 (\tau) \, C (\tau) = 0 \quad & \Longrightarrow & \quad   f_2 (\tau) \propto C(\tau), \nonumber\\
 & \Longrightarrow &  \quad  f_2 (\tau)  = \kappa_2 \, C(\tau),
\end{eqnarray} 
where $\kappa_1$ and $\kappa_2$ are the proportionality  constants. To determine the value of these constants,  we use the
generalization of  BRST invariant restriction $ s_{b} (\dot \varphi  - z) = 0$ as 
\begin{eqnarray} 
\dot\Theta (\tau, \bar\vartheta)  - {\cal Z} (\tau, \bar\vartheta)  = \dot\varphi  (\tau) - z (\tau) \quad \Longrightarrow  \quad  \kappa_1 = \kappa_2.
\end{eqnarray}  
Finally, to determine the value of constants,  we generalize the BRST 
invariant restrictions $s_{b} (\dot {\cal B} \, z + i\, \dot {\bar C}\,\dot C) = 0$
and $s_{b} ({\cal B} \,\varphi + i\,\bar C\, C) = 0$ onto $(1, 1)$-dimensional supersub-manifold as:
\begin{eqnarray} 
 {\dot {\tilde {\cal B}}} ^{(b)} (\tau, \bar\vartheta)  \,{\cal Z} (\tau, \bar\vartheta)  
+ i\,\dot {\bar F} (\tau, \bar\vartheta) \, \dot F ^{(b)} (\tau, \bar\vartheta) 
= \dot  {\cal B} (\tau) \, z (\tau) + i\, \dot {\bar C} (\tau)\,\dot C (\tau) \; \;\Longrightarrow \;\dot {b}_2 (\tau)
 = \kappa_1 \dot  {\cal B} (\tau), \nonumber\\ 
 {\tilde  {\cal B}}^{(b)}(\tau, \bar\vartheta)  \,\Theta  (\tau, \bar\vartheta)  + i\,{\bar F} (\tau, \bar\vartheta) \,  F ^{(b)} (\tau, \bar\vartheta) 
=  {\cal B} (\tau) \, \varphi  (\tau) + i\,  {\bar C} (\tau)\, C (\tau) \; \Longrightarrow  {b}_2 (\tau)
 = \kappa_2 \, {\cal B} (\tau).
\end{eqnarray}
Using the results obtained in Eq. (22) and Eq. (23), it is clear that $\kappa_1 = \kappa_2 = 1$. 
Therefore, we obtain  the value of derived variables  as: $f_1 (\tau) = \dot C (\tau), \; f_2 (\tau) 
= C (\tau), \;  b_2 (\tau) =   {\cal B} (\tau)$, thus, 
we get the following expansions for   {anti-chiral}   supervariables: 
\begin{eqnarray}
&&z (\tau)\; \longrightarrow {\cal Z} ^{(b)} (\tau, \bar\vartheta)  = z (\tau) + \bar\vartheta\, [\dot C (\tau)] 
\; \;\,\; \equiv\; z (\tau) + \bar\vartheta\, [s_b\, z (\tau)] ,\nonumber\\
&& \varphi  (\tau) \;\longrightarrow \Theta   ^{(b)}(\tau, \bar\vartheta)  = \varphi  (\tau) + \bar\vartheta\,[C(\tau)]
 \; \;\,\equiv\; \varphi  (\tau) + \bar\vartheta\,[s_{b} \,\varphi  (\tau)] , \nonumber\\
&& \bar C (\tau) \longrightarrow \bar F ^{(b)} (\tau, \bar\vartheta)  = \bar C (\tau) + \bar\vartheta\,[i\, B(\tau)] 
\;\equiv \;\bar C (\tau) + \bar\vartheta\,[s_{b} \,\bar C (\tau)].
\end{eqnarray}
Thus, in view of the above Eq. (24), we have a connection between the BRST symmetry transformation $(s_b)$
and partial derivative  $(\partial_{\bar\vartheta})$ on the   {\it anti-chiral}   supersub-manifold defined by the mapping: 
$s_b\longleftrightarrow \partial_{\bar\vartheta}$ (see, e.g. [11-13] for details). To be more clear, 
 the BRST transformation of any generic variable $\psi(\tau)$ is equal to the translation of the corresponding   {\it anti-chiral}   supervariable 
$\Psi^{(b)}(\tau, \bar\vartheta) $ along the $\bar \vartheta$-direction. Mathematically, it can be represented as 
$s_{b} \psi(\tau) = \frac{\partial}{\partial {\bar \vartheta}} \Psi^{(b)}(\tau, \bar \vartheta) = \partial_{\bar \vartheta} 
\Psi^{(b)}(\tau, \bar \vartheta)$. In other words, one can say, the coefficient of $\bar \vartheta$ in the 
expansion of an   {\it anti-chiral}   supervariable is simply   the quantum BRST  symmetry transformation of the 
corresponding variable.

We are now in the stage  to derive the quantum  anti-BRST symmetry transformations using   {\it chiral}     supervariable approach.
In this context, we use the   {\it chiral}     super expansions of the   {\it chiral}     supervariables where we generalize 
(0 + 1)-dimensional variables onto the (1, 1)-dimensional supersub-manifold of the suitably chosen most 
 {\it common}  (1, 2)-dimensional  supermanifold. The   {\it chiral}     super expansions of the ordinary variables are as follows   
\begin{eqnarray}
&&z (\tau)\; \longrightarrow \;  {\cal Z} (\tau, \vartheta)   = z (\tau) + \vartheta\,\bar f_1 (\tau),\nonumber\\
&& \varphi  (\tau) \;\longrightarrow \; \Theta  (\tau, \vartheta)   = \varphi  (\tau) + \vartheta\,\bar f_2 (\tau), \nonumber\\
&&C(\tau) \longrightarrow \; F (\tau, \vartheta)   = C(\tau) + i\,\vartheta\,\bar b_1 (\tau),\nonumber\\
&& \bar C (\tau) \longrightarrow \; \bar F (\tau, \vartheta)   = \bar C (\tau) + i\,\vartheta\,\bar b_2 (\tau), \nonumber\\
&& r (\tau) \;\longrightarrow \; R (\tau, \vartheta)   = r (\tau) + \vartheta\,\bar f_3 (\tau), \nonumber\\
&& p_r (\tau) \longrightarrow \; P_r (\tau, \vartheta)   = p_r (\tau)  + \vartheta\,\bar f_4 (\tau), \nonumber\\
&& p_\varphi  (\tau) \longrightarrow \; P_\varphi  (\tau, \vartheta)   =  p_\varphi  (\tau) + \vartheta\,\bar f_5 (\tau), \nonumber\\ 
&&  {\cal B} (\tau) \;\longrightarrow \;  {\tilde  {\cal B}} (\tau, \vartheta)   =   {\cal B} (\tau) + \vartheta\,\bar f_6 (\tau), 
\end{eqnarray}
where  derived variables $\bar b_1, \bar b_2$ are the bosonic and
 derived variables  $\bar f_1, \bar f_2, \bar f_3, \bar f_4, \bar f_5, \bar f_6$ are fermionic in nature. 
The anti-BRST invariant restrictions   also {\it must} remain independent of the Grassmannian variable ($\vartheta$) when they 
are generalized onto the (1, 1)-dimensional   {\it chiral}     supersub-manifold. The anti-BRST invariant restrictions are 
given  as
\begin{eqnarray}
&& s_{ab} (r, p_r, p_\varphi , {\cal B},   \bar C) = 0, \qquad s_{ab} (z\, \dot {\bar C}) = 0,\qquad s_{ab} (\varphi \, \bar C) = 0,\nonumber\\
&& s_{ab} (\dot {\cal B} \, z + i\, \dot {\bar C}\,\dot C) = 0, \quad s_{ab} (\dot\varphi  - z) = 0, 
\quad s_{ab} ({\cal B} \,\varphi  + i\,\bar C\, C) = 0.
\end{eqnarray}
As the physical quantities  remain independent of the Grassmannian variable $\vartheta$ which imply  that
the anti-BRST invariant restrictions can be generalized onto the (1, 1)-dimensional supersub-manifold
of the  most {\it common} (1, 2)-dimensional supermanifold as follows  
\begin{eqnarray} 
&& R (\tau, \vartheta)   = r(\tau), \; P_r (\tau, \vartheta)   = p_r (\tau), \;  P_\varphi  (\tau, \vartheta)  
 = p_\varphi  (\tau),\;{\tilde  {\cal B}
} (\tau, \vartheta)   = {\cal B} (\tau), \;  \bar F (\tau, \vartheta)   = \bar C(\tau), \nonumber\\
 &&{\cal Z} (\tau, \vartheta)  \,\dot {\bar F} (\tau, \vartheta)   = z (\tau)\, \dot {\bar C}(\tau), 
\quad \Theta  (\tau, \vartheta)  \, \bar F (\tau, \vartheta)   
= \varphi  (\tau)\, \bar C(\tau), \nonumber\\
&&{\dot {\tilde  {\cal B}}} (\tau, \vartheta)   \,{\cal Z} (\tau, \vartheta)   + i\,\dot {\bar F} (\tau, \vartheta)  \, \dot F (\tau, \vartheta)  
= \dot  {\cal B} (\tau) \, z (\tau) + i\, \dot {\bar C} (\tau)\,\dot C (\tau), \nonumber\\
&& \dot\Theta (\tau, \vartheta)   - {\cal Z} (\tau, \vartheta)   = \dot\varphi  (\tau) - z (\tau),\nonumber\\ 
&&{\tilde  {\cal B}} (\tau, \vartheta)   \,\Theta  (\tau, \vartheta)   + i\,{\bar F} (\tau, \vartheta)  \,  F (\tau, \vartheta)  
=  {\cal B} (\tau) \, \varphi  (\tau) + i\,  {\bar C} (\tau)\, C (\tau). 
\end{eqnarray}
The  above generalizations of the  anti-BRST invariant  restrictions [Eq. (26)] lead to the derivation of
 the   {\it chiral}     derived 
variables in terms  of the {\it auxiliary} and {\it basic} variables present in the Lagrangian $L$, namely;
\begin{eqnarray}
 \bar b_2  = 0, \;\; \bar f_3 = 0, \;\; \bar f_4 = 0, \;\;  \bar f_5 = 0, \;\; \bar f_6 = 0, \;\;
  \bar f_1 = \dot {\bar C}, \;\; \bar f_2 = \bar C, \;\;  \bar b_1 =  -\,{\cal B} .
\end{eqnarray}
The above values of   {\it chiral}     derived variables are derived in a similar fashion as the   {\it anti-chiral}   derived variables 
are  derived. After the substitution of the above derived variables into the   {\it chiral}     super expansions (25), we get the following 
expressions for the   {\it chiral}     supervariables onto  (1, 1)-dimensional supersub-manifold as 
\begin{eqnarray}
&& z (\tau) \;\,\longrightarrow \;{\cal Z} ^{(ab)} (\tau, \vartheta)   = z (\tau) + \vartheta\,(\dot {\bar C})
\,\;\quad  \equiv \;\; z (\tau) + \vartheta\, [s_{ab}\, z (\tau)],\nonumber\\
&& \varphi  (\tau) \;\, \longrightarrow \; \Theta  ^{(ab)} (\tau, \vartheta)   = \varphi  (\tau) + \vartheta\,(\bar C) 
\,\quad \equiv \;\; \varphi  (\tau)+ \vartheta\,[s_{ab} \varphi  (\tau)], \nonumber\\
&& C(\tau) \longrightarrow \; F ^{(ab)} (\tau, \vartheta)   = C(\tau) + \vartheta\, (- i{\cal B} ) 
\;\equiv \;\; C(\tau) + \vartheta\, [s_{ab}\, C (\tau)],\nonumber\\
&& \bar C (\tau) \longrightarrow \; \bar F ^{(ab)} (\tau, \vartheta)   = \bar C (\tau) + \vartheta\,(0) 
\;\quad  \;\equiv \;\; \bar C (\tau) + \vartheta\,[s_{ab}\bar C (\tau)]  , \nonumber\\
&& r (\tau) \;\,\longrightarrow \;  R ^{(ab)} (\tau, \vartheta)   = r (\tau) + \vartheta\,(0) 
\;\,\quad \;\equiv \;\; r (\tau) + \vartheta\,[s_{ab}\,r (\tau)] , \nonumber\\
&& p_r (\tau) \longrightarrow \; P_r ^{(ab)}(\tau, \vartheta)   = p_r (\tau) + \vartheta\,(0) 
\;\quad   \equiv \;\; p_r (\tau)  + \vartheta\,[s_{ab}\, p_r (\tau)] , \nonumber\\
&& p_\varphi  (\tau) \longrightarrow \; P_\varphi  ^{(ab)} (\tau, \vartheta)   =  p_\varphi  (\tau) + \vartheta\,(0) 
\,\quad\equiv  \;\;  p_\varphi  (\tau) + \vartheta\,[s_{ab}\, p_\varphi  (\tau)] , \nonumber\\ 
&&  {\cal B} (\tau) \; \longrightarrow \; {\tilde  {\cal B}} ^{(ab)} (\tau, \vartheta)   =   {\cal B} (\tau) + \vartheta\, (0) 
\;\;\quad   \equiv \;\; {\cal B} (\tau) + \vartheta\,[s_{ab}\, {\cal B} (\tau)]. 
\end{eqnarray}
Here, the coefficients of $\vartheta$ are simply  the anti-BRST symmetry transformations (see, e.g. [11-13] for detail). 
In fact, the anti-BRST transformation of any {\it generic} variable $\psi (\tau)$ is simply  the translation of the 
corresponding   {\it chiral}     supervariable $\Psi^{(ab)}(\tau, \vartheta)  $ along the $\vartheta$-direction. Mathematically, 
this statement can be corroborated as 
$s_{ab} \psi(\tau) = \frac{\partial}{\partial \vartheta} \Psi^{(ab)}(\tau, \vartheta)   = \partial_\vartheta \Psi^{(ab)} (\tau, \vartheta)  $.
Thus, it is clear that there is a mapping  between the quantum anti-BRST symmetry transformation $(s_{ab})$
and the Grassmannian   partial derivative  $(\partial_{\vartheta})$ defined on the {\it chiral} supersub-manifold with the mapping: 
$s_{ab} \longleftrightarrow  \partial_{\vartheta}$.\\

\section{Nilpotent (Anti-)co-BRST Symmetry Transformations: (Anti-)Chiral Supervariable Approach}

In this section, we derive the nilpotent (anti-)co-BRST symmetry transformations using the  
(anti-)chiral supervariable approach (ACSA) where we use the expansions of the (anti-)chiral
supervariables and the (anti-)co-BRST invariant restrictions.
Toward this goal in our mind, first of all, we determine  the co-BRST symmetries  by exploiting the 
{\it chiral} super expansions given in Eq. (25) and the co-BRST invariant restrictions. The co-BRST invariant restrictions are given as: 
\begin{eqnarray}
&& s_{d} (r, p_r, p_\varphi , {\cal B}, \bar C) = 0, \quad\; s_{d} (z\, {\bar C}) = 0,\quad\; s_{d} (\varphi \, \dot {\bar C})
 = 0,\nonumber\\
&&s_{d} (\varphi \,\dot  p_\varphi  + i\, \dot {\bar C}\,\dot C) = 0, \quad s_{d} (z\,  p_\varphi  - i\,\bar C\, C) = 0, 
 \quad  s_{d} (\varphi  + \dot z) = 0.
\end{eqnarray}
According to the {\it basic} rules  of ACSA, the above co-BRST invariant restrictions
can be generalized onto the (1, 1)-dimensional supersub-manifold [of the suitably chosen most {\it common}  (1, 2)-dimensional 
supermanifold] as:  
\begin{eqnarray} 
&& R (\tau, \vartheta)   = r(\tau), \; P_r (\tau, \vartheta)   = p_r (\tau), \;  P_\varphi  (\tau, \vartheta)   = p_\varphi  (\tau),\;
 {\tilde  {\cal B}} (\tau, \vartheta)   = {\cal B} (\tau), \; \bar F (\tau, \vartheta)   = \bar C(\tau), \nonumber\\
 &&{\cal Z} (\tau, \vartheta)  \,{\bar F} (\tau, \vartheta)   = z (\tau)\, {\bar C}(\tau), \quad \Theta  (\tau, \vartheta)  \,
 \dot {\bar F} (\tau, \vartheta)   = \varphi  (\tau)\, \dot {\bar C}(\tau), \nonumber\\
 &&\Theta (\tau, \vartheta)  \, \dot P_\varphi  (\tau, \vartheta)   \, + i\,\dot {\bar F} (\tau, \vartheta)  \,  \dot F (\tau, \vartheta)  
= \varphi  (\tau)\,\dot  p_\varphi  (\tau) + i\, \dot {\bar C} (\tau)\,\dot C (\tau),\nonumber\\
&&{\cal Z} (\tau, \vartheta)  \,P_\varphi  (\tau, \vartheta)   -  i\,{\bar F} (\tau, \vartheta)  \,  F (\tau, \vartheta)  
= z (\tau) \,p_\varphi  (\tau) - i\, {\bar C} (\tau)\,C (\tau), \nonumber\\
&& \Theta (\tau, \vartheta)   +\dot  {\cal Z} (\tau, \vartheta)   = \varphi  (\tau) + \dot z (\tau).
\end{eqnarray}
At this stage, we determine  the value of derived variables of Eq. (25) using the above generalizations
of the co-BRST invariant restrictions. To derive the value of the derived variables, first of all, 
we use the first line entry of  Eq. (31) where the trivial co-BRST invariant quantities are generalized
which implies the following relationships: 
 \begin{eqnarray} 
&& P_\varphi  (\tau, \vartheta) = p_\varphi  (\tau) \Longrightarrow \;  \bar f_5 = 0, \qquad \,\;\;\;  {\tilde  {\cal B}} (\tau, \vartheta)  
 = {\cal B} (\tau) \Longrightarrow \bar f_6 = 0, \nonumber\\
&& R (\tau, \vartheta)   \; = r(\tau)\; \Longrightarrow \; \bar f_3 = 0, \qquad \quad \bar F (\tau, \vartheta)  
 = \bar C(\tau) \Longrightarrow \bar b_2 = 0, \nonumber\\
&& P_r (\tau, \vartheta)   = p_r (\tau) \Longrightarrow \; \bar f_4 = 0. 
\end{eqnarray}
After substituting the above value of derived variables into the expressions of the    {\it chiral}     
super expansions [Eq. (25)], we obtain  the following   {\it chiral}     super expansions:
\begin{eqnarray}
&&\bar C(\tau) \, \longrightarrow \; \bar F ^{(d)} (\tau, \vartheta)   = \bar C(\tau) + \vartheta\,(0) \;\equiv \;  \bar C(\tau) 
+ \vartheta\,[s_d\, \bar C (\tau)] ,\nonumber\\
&& r (\tau) \;\, \longrightarrow \; R ^{(d)} (\tau, \vartheta)   = r (\tau) + \vartheta\,(0) \; \;\equiv \; r (\tau) + \vartheta\,[s_d \,r (\tau)], \nonumber\\
&& p_r (\tau) \longrightarrow \; P_r  ^{(d)} (\tau, \vartheta)   = p_r (\tau)  + \vartheta\, (0) \;\equiv \;  p_r (\tau)  + \vartheta\, [s_d\, p_r (\tau)], \nonumber\\
&& p_\varphi  (\tau) \longrightarrow \; P_\varphi  ^{(d)} (\tau, \vartheta)   =  p_\varphi  (\tau) + \vartheta\, (0)
 \equiv \; p_\varphi  (\tau) + \vartheta\, [s_d\, p_\varphi  (\tau)], \nonumber\\ 
&&  {\cal B} (\tau) \; \longrightarrow \; {\tilde  {\cal B}}  ^{(d)} (\tau, \vartheta)   =   {\cal B} (\tau) + \vartheta\,(0) 
\,\; \equiv \;  {\cal B} (\tau) + \vartheta\,[s_d\, {\cal B} (\tau)], 
\end{eqnarray}
where superscript $(d)$ on the   {\it chiral}     supervariables  denotes the supervariables obtained after the 
application of the co-BRST (i.e. dual-BRST) invariant restrictions. 
For the non-trivial case, first of all, we generalize the co-BRST invariant restriction 
$s_{d} (z\, {\bar C}) = 0$ and $s_{d} (\varphi \, \dot {\bar C}) = 0$  onto (1, 1)-dimensional supersub-manifold as   
 \begin{eqnarray} 
{\cal Z} (\tau, \vartheta)  \,{\bar F}^{(d)} (\tau, \vartheta)   =  z (\tau)\, {\bar C}(\tau), \qquad \Theta (\tau, \vartheta)  \,
 \dot {\bar F} ^{(d)} (\tau, \vartheta)   = \varphi  (\tau)\, \dot {\bar C}(\tau),
\end{eqnarray}
which lead to the following relationships for the derived variables
\begin{eqnarray} 
\bar f_1 (\tau)\,\bar C(\tau) = 0 \quad & \Longrightarrow & \quad \bar f_1 (\tau) \propto \bar C(\tau), \nonumber\\
& \Longrightarrow & \quad \bar f_1 (\tau)  =  -\,\bar \kappa _1 \bar C(\tau), \nonumber\\
\bar f_2 (\tau) \, \dot {\bar C} (\tau) = 0 \quad & \Longrightarrow & \quad   \bar f_2 (\tau) \propto \dot {\bar C}(\tau), \nonumber\\
 & \Longrightarrow &  \quad  \bar f_2 (\tau)  = \bar \kappa_2  \dot {\bar C}(\tau),
\end{eqnarray}
where $\bar\kappa_1$ and $\bar\kappa_2$ are the proportionality constants.
To determine  the value of  these constants, we further use the generalizations
of the co-BRST invariant restrictions $s_{d} (\varphi  + \dot z)  = 0, \; s_{d} (\varphi \,\dot  p_\varphi  
+ i\, \dot {\bar C}\,\dot C) = 0$ and   $s_{d} (z\,  p_\varphi - i\,\bar C\, C) = 0$ as:
\begin{eqnarray} 
\Theta  (\tau, \vartheta)   + \dot {\cal Z} (\tau, \vartheta)   = \varphi  (\tau) + \dot z (\tau) \quad & \Longrightarrow &   \bar\kappa_1 
= \bar\kappa_2, \nonumber\\ 
\Theta (\tau, \vartheta)  \,\dot P_\varphi  ^{(d)} (\tau, \vartheta)    + i\,\dot {\bar F} ^{(d)} (\tau, \vartheta)  \, \dot F (\tau, \vartheta)  
& = & \varphi  (\tau) \,\dot p_\varphi  (\tau) \, + i\, \dot {\bar C} (\tau)\,\dot C (\tau) \nonumber\\
  & \Longrightarrow & \;\;\dot {\bar b}_1 (\tau)
= -\,\bar\kappa_2 \, \dot p_\varphi  (\tau), \nonumber\\
{\cal Z} (\tau, \vartheta)  \,P_\varphi  ^{(d)}(\tau, \vartheta)    - i\,{\bar F} ^{(d)} (\tau, \vartheta)  \,  F (\tau, \vartheta)  
& = & z (\tau)\, p_\varphi  (\tau)\,  - i\,  {\bar C} (\tau)\, C (\tau)\nonumber\\
   & \Longrightarrow & \bar b_1 (\tau) = -\,\bar\kappa_1 \, p_\varphi  (\tau). 
\end{eqnarray}
The results of the above three relations in Eq. (36) imply  that   proportionality constant are  equal 
(i.e. $\bar \kappa_1 =  \bar\kappa_2$) and their values are equal to minus  one (i.e $\bar\kappa_1 =  \;\bar\kappa_2 = -\, 1$). Therefore, we get the value
 of the derived variables as: $\bar f_1 = \bar C, \; \bar f_2 = -\,\dot {\bar C},\;\bar b_1 =  p_\varphi $.
 As a result, we have the following   {\it chiral}     super expansions of the ordinary variables:  
 \begin{eqnarray}
&&z (\tau) \; \longrightarrow \; {\cal Z} ^{(d)} (\tau, \vartheta)   = z (\tau) + \vartheta\, [\bar C (\tau)] 
\;\; \;\;\equiv z (\tau) + \vartheta\, [s_d \,z (\tau)] ,\nonumber\\
&& \varphi  (\tau) \; \longrightarrow \; \Theta   ^{(d)}(\tau, \vartheta)   = \varphi  (\tau) + \vartheta\,[-\,\dot {\bar C}(\tau)]
 \equiv \varphi  (\tau) + \vartheta\,[s_d \,\varphi  (\tau)] , \nonumber\\
&& C (\tau) \longrightarrow \; \bar F ^{(d)} (\tau, \vartheta)   = C (\tau) + \vartheta\,[i\, p_\varphi  (\tau)] 
\;\equiv  C (\tau) + \vartheta\,[s_d \, C (\tau)].
\end{eqnarray}
where superscript $(d)$ on the  supervariables denotes the same meaning as in Eq. (33). From the above equations, 
it is clear that the translation of any generic   {\it chiral}     supervariable $\Psi^{(d)}(\tau, \vartheta)  $ along the $\vartheta$-direction generates co-BRST symmetry transformation ($s_d$) of the corresponding variable $\psi(\tau)$. Mathematically, we can express this statements as 
$s_d\psi(\tau) = \partial_\vartheta \Psi^{(d)}(\tau, \vartheta)  $.
In other words, the coefficients of the $\vartheta$ are simply the co-BRST symmetry transformations. Thus, the 
co-BRST symmetry transformation $(s_d)$ is connected with the Grassmannian derivative
 $\partial_\vartheta$ (i.e. $s_d \longleftrightarrow  \partial_\vartheta$) [11-13].

Now for the derivation of anti-co-BRST symmetry transformations, we use the   {\it anti-chiral}   
super expansions of the supervariables [Eq. (15)] and  
anti-co-BRST invariant restrictions. The  anti-co-BRST invariant restrictions are listed as 
\begin{eqnarray}
&& s_{ad} (r, p_r, p_\varphi , {\cal B},   C) = 0, \qquad\; s_{ad} (z\, {C}) = 0,\qquad\; s_{ad} (\varphi \, \dot {C}) = 0,\nonumber\\
&& s_{ad} (z\,  p_\varphi - i\,\bar C\, C) = 0, \quad s_{ad} (\varphi \,\dot  p_\varphi  
 + i\, \dot {\bar C}\,\dot C) = 0, \quad s_{ad} (\dot z +  \varphi ) = 0.
\end{eqnarray}
The generalization of these  anti-co-BRST invariant restrictions  onto the (1, 1)-dimensional supersub-manifold
(of the most {\it common}  (1, 2)-dimensional supermanifold) are as follows: 
\begin{eqnarray*} 
&& R (\tau, \bar\vartheta)  = r(\tau), \; P_r (\tau, \bar\vartheta)  = p_r (\tau), \;  P_\varphi 
 (\tau, \bar\vartheta)  = p_\varphi  (\tau),\;
 {\tilde  {\cal B}} (\tau, \bar\vartheta)  =  {\cal B}  (\tau), \;  F (\tau, \bar\vartheta)  = C(\tau), \nonumber\\
 \end{eqnarray*}
 \begin{eqnarray}
 &&{\cal Z} (\tau, \bar\vartheta) \, F (\tau, \bar\vartheta)  = z (\tau)\, C(\tau), \quad \Theta  (\tau, \bar\vartheta) \, \dot F (\tau, \bar\vartheta)  
= \varphi  (\tau)\, \dot C(\tau), \nonumber\\
&&{\cal Z} (\tau, \bar\vartheta) \,P_\varphi   (\tau, \bar\vartheta)   -  i\,{\bar F} (\tau, \bar\vartheta) \, F (\tau, \bar\vartheta) 
= z (\tau)\,  p_\varphi  (\tau)  - i\,  {\bar C} (\tau)\, C (\tau), \nonumber\\
 &&\Theta  (\tau, \bar\vartheta) \, \dot P_\varphi  (\tau, \bar\vartheta)  \, + i\,\dot {\bar F} (\tau, \bar\vartheta) \,  \dot F (\tau, \bar\vartheta) 
= \varphi  (\tau) \, p_\varphi  (\tau) + i\,  \dot {\bar C} (\tau)\, \dot C (\tau), \nonumber\\
&& \dot {\cal Z} (\tau, \bar\vartheta)  +   \Theta (\tau, \bar\vartheta)  =  \dot z (\tau)   + \varphi  (\tau).
\end{eqnarray} 
The above generalizations of the anti-co-BRST invariant restrictions, finally,
 lead to the derivation of the derived variables as follows: 
\begin{eqnarray}
 b_1  = 0, \;\;  b_2 = -\, p_\varphi , \;\; f_1 = {C}, \;\;  f_2
 =  -\,\dot C, \;\;   f_3 = 0, \;\; f_4 = 0, \;\;  f_5 = 0, \;\; f_6 = 0.
\end{eqnarray}
Thus, we have determined all the derived variables using the same technique as we have used in the derivation of the 
co-BRST symmetry transformations. Finally, after substituting the value of derived variables  into Eq. (15),
we obtain the following expressions for the   {\it anti-chiral}   expansions of the supervariables, namely;
\begin{eqnarray}
&& z (\tau) \; \longrightarrow \; {\cal Z} ^{(ad)} (\tau, \bar\vartheta)  = z (\tau) + \bar\vartheta\,(C)  \quad\;\;\;  \equiv \;\;  z (\tau)
 + \bar\vartheta\, [s_{ad}\, z (\tau)],\nonumber\\
&& \varphi  (\tau) \; \longrightarrow \; \Theta  ^{(ad)} (\tau, \bar\vartheta)  = \varphi  (\tau) + 
\bar\vartheta\,(-\,\dot C) \;\; \equiv \;\;  \varphi  (\tau) + \bar\vartheta\,[s_{ad}\, \varphi  (\tau)], \nonumber\\
&& C(\tau) \longrightarrow \; F ^{(ad)} (\tau, \bar\vartheta)  = C(\tau) + \bar\vartheta\, (0) \qquad \equiv \;\; C(\tau) + \bar\vartheta\, [s_{ad}\, C (\tau)],\nonumber\\
&& \bar C (\tau) \longrightarrow \;  \bar F ^{(ad)} (\tau, \bar\vartheta)  = \bar C (\tau) + \bar\vartheta\,(-\,i\,p_\varphi ) \equiv \;\;  \bar C (\tau) 
+ \bar\vartheta\,[s_{ad}\, \bar C (\tau)], \nonumber\\
&& r (\tau) \; \longrightarrow \; R ^{(ad)} (\tau, \bar\vartheta)  = r (\tau) + \bar\vartheta\,(0) \qquad \;\equiv \;\; r (\tau) + \bar\vartheta\,[s_{ad}\, r (\tau)], \nonumber\\
&& p_r (\tau) \longrightarrow \; P_r ^{(ad)}(\tau, \bar\vartheta)  = p_r (\tau)  + \bar\vartheta\,(0) \;\;\,\quad \equiv \;\;  p_r (\tau)  
+ \bar\vartheta\,[s_{ad}\, p_r (\tau)] , \nonumber\\
&& p_\varphi  (\tau) \longrightarrow \; P_\varphi  ^{(ad)} (\tau, \bar\vartheta)  =  p_\varphi  (\tau) + \bar\vartheta\,(0) \quad\;\; \equiv \;\; p_\varphi  (\tau)
+ \bar\vartheta\,[s_{ad}\, p_\varphi  (\tau)] , \nonumber\\ 
&&  {\cal B} (\tau) \, \longrightarrow \; {\tilde  {\cal B}} ^{(ad)} (\tau, \bar\vartheta)  =   {\cal B} (\tau)
 + \bar\vartheta\, (0) \qquad \equiv \;\; {\cal B} (\tau) + \bar\vartheta\,[s_{ad}\, {\cal B} (\tau)], 
\end{eqnarray}
where superscript $(ad)$ on the   {\it anti-chiral}   supervariables denote the fact that   {\it anti-chiral}   supervariables are obtained after the
application of quantum anti-co-BRST invariant restrictions [Eq. (38)]. 
Here, it is clear, the coefficient of Grassmannian variable $\bar\vartheta$ is simply  the quantum anti-co-BRST 
symmetry transformation ($s_{ad}$). To be more clear, the anti-co-BRST symmetry transformation $(s_{ad})$ of any generic 
variable $\psi(\tau)$ is equal to the translation of the corresponding   {\it anti-chiral}   supervariable 
$\Psi^{(ad)}(\tau, \bar \vartheta)$ along the $\bar \vartheta$-direction i.e.
$s_{ad} \psi(\tau) = \partial_{\bar \vartheta} \Psi^{(ad)}(\tau, \bar \vartheta)$. This implies that the 
anti-co-BRST symmetry ($s_{ad}$) is connected with the Grassmannian translation generator 
($\partial_{\bar\vartheta}$) as $s_{ad} \longleftrightarrow \partial_{\bar\vartheta}$ [11-13].\\

\noindent
\section{Nilpotency and Absolute Anti-Commutativity  of the Noether Conserved  Charges: ACSA}

In this section, we deduce   the  nilpotency and absolute anti-commutativity properties 
of the conserved (anti-)BRST  along with   (anti-)co-BRST  charges in the language of ACSA. 
For this, first of all, we show the nilpotency of the (anti-)BRST together with 
(anti-)co-BRST conserved charges. It is straightforward to express the expressions of the  (anti-)BRST together with  (anti-)co-BRST
charges in terms of the (anti-)chiral supervariables  and partial derivatives  $(\partial_{\bar\vartheta}, \partial_\vartheta)$ 
with an equivalent integral form as follows  
\begin{eqnarray}
{\cal Q}_{b} & = &  \frac {\partial}{\partial \bar \vartheta} \Big[i\, \dot {\bar F} ^ {(b)} (\tau, \bar \vartheta)\, F ^ {(b)} (\tau, \bar \vartheta) 
- i\, \bar F ^ {(b)} (\tau, \bar \vartheta) \, \dot F ^ {(b)} (\tau, \bar \vartheta) \Big] \nonumber\\
& \equiv &  \int \,d \bar\vartheta \Big[i\, \dot {\bar F} ^ {(b)} (\tau, \bar \vartheta)\, F ^ {(b)} (\tau, \bar \vartheta) 
- i\, \bar F ^ {(b)} (\tau, \bar \vartheta) \, \dot F ^ {(b)} (\tau, \bar \vartheta)\Big],\nonumber\\
{\cal Q}_{ab} & = & \,\frac {\partial}{\partial \vartheta}\, \Big[i\, \bar F ^ {(ab)} (\tau, \vartheta)   \, \dot F ^ {(ab)} (\tau,  \vartheta)
 - i\, \dot {\bar F} ^ {(ab)} (\tau,  \vartheta)\, F ^ {(ab)} (\tau, \vartheta)  \Big] \nonumber\\
&\equiv & \,\int \, d \vartheta\, \Big[i\, \bar F ^ {(ab)} (\tau, \vartheta)   \, \dot F ^ {(ab)} (\tau, \vartheta)   
- i\, \dot {\bar F} ^ {(ab)} (\tau, \vartheta)  \, F  ^ {(ab)} (\tau, \vartheta)  \Big], 
\end{eqnarray}
\begin{eqnarray}
{\cal Q}_d &=& \frac {\partial}{\partial \vartheta}\, \Big[i\,\bar F ^ {(d)} (\tau, \vartheta)   \, \dot F ^ {(d)} (\tau, \vartheta)   
- i\, \dot {\bar F} ^ {(d)} (\tau, \vartheta)  \, F ^ {(d)} (\tau, \vartheta)   \Big]\nonumber\\ 
&\equiv & \int \, d \vartheta \, \Big[i\,\bar F ^ {(d)} (\tau, \vartheta)   \, \dot F  ^ {(d)} (\tau, \vartheta)   
- i\, \dot {\bar F} ^ {(d)} (\tau, \vartheta)  \, F ^ {(d)}  (\tau, \vartheta)   \Big], \nonumber\\
{\cal Q}_{ad} &=& \frac {\partial}{\partial \bar \vartheta}\, \Big[i\, \dot {\bar F} ^ {(ad)}(\tau, \bar\vartheta) \, F ^ {(ad)}(\tau, \bar\vartheta)  
- i\,\bar F ^ {(ad)} (\tau, \bar\vartheta)  \, \dot F  ^ {(ad)}(\tau, \bar\vartheta)   \Big]\nonumber\\ 
&\equiv & \int \, d \bar\vartheta \, \Big[i\, \dot {\bar F} ^ {(ad)}(\tau, \bar\vartheta) \, F  ^ {(ad)}(\tau, \bar\vartheta)  
- i\,\bar F  ^ {(ad)}(\tau, \bar\vartheta)  \, \dot F  ^ {(ad)}(\tau, \bar\vartheta)   \Big],
\end{eqnarray}
where the superscripts $(b)$ and $(ab)$ stand for the   {\it anti-chiral}   and   {\it chiral}     supervariables  that have been obtained after
the application of the BRST and anti-BRST invariant restrictions,  respectively. The superscripts  
$(d)$ and  $(ad)$ show the   {\it chiral}     and    {\it anti-chiral}   supervariables that is  obtained after the application of 
co-BRST and anti-co-BRST  invariant restrictions, respectively.
It is clear that the nilpotency ($\partial_{\bar\vartheta}^ 2 = 0, \;\partial_{\vartheta}^ 2 = 0$) of the translational generators
$(\partial_{\bar\vartheta}, \; \partial_{\vartheta})$ implies that
\begin{eqnarray}
&&\partial_{\bar\vartheta}\; {\cal Q}_{b}  = 0 
\quad\;\,\Longleftrightarrow \quad s_b\; {\cal Q}_{b}  \quad \; = \; -\;i\;{\{{\cal Q}_{b},   {\cal Q}_b}\} = 0,\nonumber\\
&&\partial_{\vartheta}\; {\cal Q}_{ab}  = 0
\quad\Longleftrightarrow \quad s_{ab}\; {\cal Q}_{ab}  \; \,\,= \; -\;i\;{\{{\cal Q}_{ab}, {\cal Q}_{ab}}\} = 0,\nonumber\\
&&\partial_{\vartheta}\; {\cal Q}_d  = 0 
\quad\;\Longleftrightarrow \quad s_d\; {\cal Q}_d  \; \quad = \; -\;i\;{\{{\cal Q}_d, {\cal Q}_d}\} = 0,\nonumber\\
&&\partial_{\bar\vartheta}\; {\cal Q}_{ad}  = 0
\quad\Longleftrightarrow \quad s_{ad}\; {\cal Q}_{ad}  \;\; = \; -\;i\;{\{{\cal Q}_{ad}, {\cal Q}_{ad}}\} = 0,
\end{eqnarray}
which show the nilpotency  [${\cal Q}_{(a)b} ^2 = {\cal Q}_{(a)d} ^2 = 0$] of the conserved charges within
 the ambit of ACSA to BRST formalism. Thus, we have shown that there is a deep connection 
between the nilpotency ($\partial_{\bar\vartheta}^ 2 = 0, \; \partial_{\vartheta}^ 2 = 0$) of the translational
generator $(\partial_{\bar\vartheta}, \; \partial_{\vartheta})$ and the nilpotency [i.e. 
${\cal Q}_{(a)b}^2 =  {\cal Q}_{(a)d}^2 = 0$] of the (anti-)BRST and (anti-)co-BRST charges [${\cal Q}_{(a)b}, \; {\cal Q}_{(a)d}$].
The above  nilpotency property can  be also captured in an ordinary space where we use the (anti-)BRST {\it exact} as well as 
(anti-)co-BRST {\it exact} 
forms of the charges, namely; 
\begin{eqnarray}
{\cal Q}_{b} &=& -\, i\,s_b\, \big(\bar C \, \dot C - \dot {\bar C}\, C\big),  \qquad {\cal Q}_{ab}  
=  +\,i \,s_{ab}\, \big(\bar C \, \dot C - \dot {\bar C}\, C\big),\nonumber\\
{\cal Q}_d &=& i\,s_d\, \big(\bar C \, \dot C - \dot {\bar C}\, C\big), \;\;\; \qquad {\cal Q}_{ad} 
= - \,i \,s_{ad}\, \big(\bar C \, \dot C - \dot {\bar C}\, C\big), 
\end{eqnarray}
the above expressions  show  the nilpotency property  of the (anti-)BRST along with  (anti-) co-BRST 
conserved charges, in a simpler way, in an ordinary space [cf. (44)].

Now, we are in a stage to show the absolute anti-commutativity of the (anti-)BRST along with   (anti-)co-BRST
charges. For this purpose, we write the charges in terms of the (anti-) chiral supervariables and 
the derivatives $(\partial_\vartheta, \; \partial_{\bar\vartheta})$ of the Grassmannian variables $(\bar\vartheta, \vartheta)$ 
\begin{eqnarray}
{\cal Q}_{b} &=&  - \,i\,\frac {\partial}{\partial \vartheta}  \,\Big[\dot F ^ {(ab)} (\tau, \vartheta)    \, F ^ {(ab)} (\tau, \vartheta)  \Big]   
   \, \equiv   - \,i\,\int \, d \vartheta  \,\Big[\dot F ^ {(ab)} (\tau, \vartheta)    \, F ^ {(ab)} (\tau, \vartheta)  \Big],\nonumber\\
{\cal Q}_{ab} &=&  \;\;\; i\,\frac {\partial}{\partial \bar \vartheta}  \,\Big[\dot {\bar F} ^ {(b)} (\tau, \bar \vartheta)  \, \bar F ^ {(b)} (\tau, \bar \vartheta)\Big] \quad  \equiv  \;\; \; i\,\int \, d \bar \vartheta  \,\Big[\dot {\bar F } ^ {(b)} (\tau, \bar\vartheta)   \, \bar F ^ {(b)} (\tau, \bar\vartheta) \Big],\nonumber\\
{\cal Q}_d &=&  \;\; \; i\, \frac {\partial}{\partial \bar \vartheta} \,\Big[\dot {\bar F} ^ {(ad)} (\tau, \bar\vartheta)  \,
 \bar F ^ {(ad)} (\tau, \bar\vartheta) \Big] \, \equiv   \; \;\; i\, \int d \bar\vartheta \,\Big[\dot {\bar F} ^ {(ad)} (\tau, \bar\vartheta) 
 \, \bar F ^ {(ad)}(\tau, \bar\vartheta) \Big],\nonumber\\
{\cal Q}_{ad} &=&  -\,i\, \frac {\partial}{\partial  \vartheta} \,\Big[\dot {F} ^ {(d)} (\tau, \vartheta)   \, F ^ {(d)} (\tau, \vartheta)  \Big] 
 \quad \equiv   -\,i\, \int d \vartheta \,\Big[\dot {F} ^ {(d)} (\tau, \vartheta)   \, F ^ {(d)} (\tau, \vartheta)  \Big], 
\end{eqnarray}
where the superscripts $(a)b$ and $(a)d$ denote the same meaning as explained earlier.
Here, it is straightforward to check  that the nilpotency ($\partial_{\bar\vartheta}^ 2 = 0, \;\partial_{\vartheta}^ 2 = 0$) of
the translational generators $(\partial_{\bar\vartheta}, \; \partial_{\vartheta})$ implies that
the following relations 
\begin{eqnarray}
&&\partial_{\vartheta}\; {\cal Q}_{b}  = 0 \; \quad\, 
\Longleftrightarrow \quad s_{ab}\; {\cal Q}_{b}  = -\,i\;{\{{\cal Q}_{b},   {\cal Q}_{ab}}\} = 0,\nonumber\\
&&\partial_{\bar\vartheta}\;{\cal Q}_{ab} = 0 \quad 
\Longleftrightarrow \quad s_b\;{\cal Q}_{ab} = - \,i\;{\{{\cal Q}_{ab}, {\cal Q}_b}\} = 0,\nonumber\\
&&\partial_{\bar\vartheta}\; {\cal Q}_d  = 0 \; \quad 
\Longleftrightarrow \quad s_{ad}\; {\cal Q}_d  = -\,i\;{\{{\cal Q}_d, {\cal Q}_{ad}}\} = 0,\nonumber\\
&&\partial_{\vartheta}\;{\cal Q}_{ad} = 0 \quad 
\Longleftrightarrow \quad s_d\;{\cal Q}_{ad} = - \,i\;{\{{\cal Q}_{ad}, {\cal Q}_d}\} = 0,
\end{eqnarray}
which show the absolute anti-commutativity property of the (anti-)BRST as well as  (anti-)co-BRST
conserved charges. The property of absolute anti-commutativity of conserved charges  can also be 
captured explicitly in an ordinary space by using the following 
(anti-)BRST {\it exact} and (anti-)co-BRST {\it exact} forms of the charges, namely;
\begin{eqnarray}
&& {\cal Q}_{b} = - \,i\, s_{ab} \,\big(\dot C \, C\big), \qquad {\cal Q}_{ab} =  +\,i\, s_{b} \,\big(\dot {\bar C} \, \bar C\big),  \nonumber\\
&& {\cal Q}_d  = i\, s_{ad} \,\big(\dot {\bar C} \, \bar C\big), \;\;\; \qquad  {\cal Q}_{ad} =  - \, i\, s_d \,\big(\dot C \, C\big). 
\end{eqnarray}\\

\noindent
\section{Invariances of Lagrangian in ACSA}

In this section, we discus the (anti-)BRST along with  (anti-)co-BRST invariances of the Lagrangian (7)
 within the scope  of ACSA to BRST formalism. 
For this purpose, foremost, we generalize the ordinary Lagrangian of (0 + 1)-dimensional onto
 the suitably chosen (1, 1)-dimensional (anti-)chiral supersub-manifold of the most {\it common}  (1, 2)-dimensional   
supermanifold. The expressions of the (anti-)chiral super  Lagrangian are
\begin{eqnarray*}
L(\tau) \longrightarrow \tilde L ^{(ac)} (\tau, \bar\vartheta)  & = & \dot r (\tau) \,p_r (\tau) + \dot \Theta   ^ {(b)}(\tau, \bar\vartheta) \, p_\varphi  (\tau) - \frac{1}{2}\, p^2_r (\tau) - \frac{1}{2 r^2}\, p^2_\varphi  (\tau)  \nonumber\\
& - &  {\cal Z} ^ {(b)}(\tau, \bar\vartheta)  \,p_\varphi  (\tau) -  V(r) +   \frac{1}{2}\,{\cal B} ^2 (\tau) 
+   {\cal B} (\tau) \,\big[\dot {\cal Z} ^ {(b)}(\tau, \bar\vartheta)    \nonumber\\
& + &  \Theta   ^ {(b)}(\tau, \bar\vartheta) \big] -  i\,\dot {\bar F}^ {(b)}(\tau, \bar\vartheta) \, \dot C (\tau)  +  i \, \bar F ^ {(b)}(\tau, \bar\vartheta)  \, C(\tau),\nonumber\\
\end{eqnarray*}
\begin{eqnarray}
L(\tau) \longrightarrow \tilde L ^{(c)} (\tau, \vartheta)   & = & \dot r (\tau) \,p_r (\tau) + \dot \Theta    ^ {(ab)}(\tau, \vartheta)  \, p_\varphi  (\tau)
 - \frac{1}{2}\, p^2_r (\tau) - \frac{1}{2 r^2}\, p^2_\varphi  (\tau)\nonumber\\
 & - & {\cal Z} ^ {(ab)}(\tau, \vartheta)   \,p_\varphi  (\tau)   -   V(r) +   \frac{1}{2}\,{\cal B}  ^2 (\tau) 
+   {\cal B} (\tau) \,\big[\dot {\cal Z} ^ {(ab)}(\tau, \vartheta)   \nonumber\\
 & + & \Theta    ^ {(ab)}(\tau, \vartheta)  \big]   -  i\,\dot {\bar C} (\tau)\, \dot F ^ {(ab)}(\tau, \vartheta)  +  i \, \bar C (\tau) \, F ^ {(ab)}(\tau, \vartheta) ,
\end{eqnarray}
where the superscripts $(ac)$ and $(c)$ on the super Lagrangians denote the   {\it anti-chiral}   and   {\it chiral}     super  Lagrangians (containing   
{\it anti-chiral} and   {\it chiral}     supervariables), respectively.   
It is evident that under  the application of translational generators $(\partial_{\bar\vartheta}, \; \partial_\vartheta)$, 
we get the (anti-) BRST invariance of Lagrangian  ($L$) with the  following results  
 \begin{eqnarray}
&& \frac {\partial}{\partial {\bar\vartheta}} \Big[\tilde L ^{(ac)} (\tau, \bar\vartheta) \Big]  = \frac {d}{d\,\tau} [ {\cal B} (\tau)\,\dot C (\tau)],\nonumber\\ 
&& \frac {\partial}{\partial {\vartheta}} \Big[\tilde L ^{(c)} (\tau, \vartheta)  \Big]  \; = \frac {d}{d\,\tau} [ {\cal B} (\tau)\,\dot {\bar C} (\tau)],
\end{eqnarray}
which imply  that generalized version of super Lagrangians remain quasi-invariant (i.e. up to a  total time derivative)
under the translational generators $(\partial_{\bar\vartheta}, \; \partial_\vartheta)$  within the scope  
 of ACSA which are consistent with Eq. (10).

Now, we capture the (anti-)co-BRST invariance of the Lagrangian (7)
within the scope  of (anti-)chiral supervariable approach. 
For this, we generalize the ordinary Lagrangian into (anti-)co-BRST super Lagrangian 
where (0 + 1)-dimensional theory is generalized onto the (1, 1)-dimensional 
(anti-)chiral supersub-manifold of the {\it common} (1, 2)-dimensional supermanifold as follows:    
\begin{eqnarray}
L(\tau) \longrightarrow \tilde L ^{(c,\, d)} (\tau, \vartheta)   & = & \dot r (\tau) \,p_r (\tau) + \dot \Theta    ^ {(d)}(\tau, \vartheta)  \, p_\varphi  (\tau)
 - \frac{1}{2}\, p^2_r (\tau) - \frac{1}{2 r^2}\, p^2_\varphi  (\tau)\nonumber\\
& - & {\cal Z} ^ {(d)}(\tau, \vartheta)   \,p_\varphi  (\tau) -  V(r)  +   \frac{1}{2}\,{\cal B}  ^2 (\tau) 
+   {\cal B} (\tau) \,\big[\dot {\cal Z} ^ {(d)}(\tau, \vartheta)  \nonumber\\
& + &  \Theta   ^ {(d)}(\tau, \vartheta)  \big]  
- i\,\dot {\bar C} (\tau)\, \dot F ^ {(d)}(\tau, \vartheta) +  i \, \bar C (\tau)\, F ^ {(d)}(\tau, \vartheta)  ,\nonumber\\
L(\tau) \longrightarrow \tilde L ^{(ac,\, ad)} (\tau, \bar\vartheta)  & = & \dot r (\tau) \,p_r (\tau) + \dot \Theta   ^ {(ad)}(\tau, \bar\vartheta) \, p_\varphi  (\tau) - \frac{1}{2}\, p^2_r (\tau) - \frac{1}{2 r^2}\, p^2_\varphi  (\tau) \nonumber\\
 & - &  {\cal Z} ^ {(ad)}(\tau, \bar\vartheta)  \,p_\varphi  (\tau) -  V(r) +   \frac{1}{2}\, {\cal B} ^2 (\tau) +   {\cal B} (\tau) \,\big[\dot {\cal Z} ^ {(ad)}(\tau, \bar\vartheta)  \nonumber\\
& + &  \Theta  ^ {(ad)}(\tau, \bar\vartheta) \big] 
 - i\,\dot {\bar F}^ {(ad)}(\tau, \bar\vartheta) \, \dot C (\tau) +  i \, \bar F ^ {(ad)}(\tau, \bar\vartheta)  \, C(\tau),
\end{eqnarray}
where the superscripts $(c,\, d)$ and  $(ac,\, ad)$ denote that the super Lagrangians (containing the 
chiral and   {\it anti-chiral}   supervariables) obtained after the application of the co-BRST and anti-co-BRST 
invariant restrictions, respectively. It is straightforward to check that
\begin{eqnarray}
&& \frac {\partial}{\partial {\vartheta}} \Big[\tilde L ^{(c, d)} (\tau, \vartheta)  \Big]  \;\; \,= \; -\,\frac {d}{d\,\tau} 
\big[p_\varphi  (\tau)\,\dot {\bar C} (\tau)\big],\nonumber\\
&& \frac {\partial}{\partial {\bar\vartheta}} \Big[\tilde L ^{(ac, ad)} (\tau, \bar\vartheta) \Big]  
= \; -\,\frac {d}{d\,\tau} \big[p_\varphi  (\tau)\,\dot C (\tau)\big],
\end{eqnarray}
which show the (anti-)co-BRST invariance of the Lagrangian $L$ within the ambit of ACSA to BRST formalism. 
At the end of this section, we have the following  concluding remarks. 
There are deep connections between the (anti-)BRST symmetries  $(s_{(a)b})$ and derivatives
 $(\partial_{\bar\vartheta}, \; \partial _\vartheta)$ of the Grassmannian variables $(\bar\vartheta, \vartheta)$ with the following mappings:
 $s_{b} \longleftrightarrow  \partial _{\bar\vartheta}$ and  $s_{ab} \longleftrightarrow \partial _{\vartheta}$. Similarly, in the case of 
(anti-)co-BRST symmetry transformations, it is clear that these symmetry transformations  are also connected with the derivatives  
$(\partial _{\bar\vartheta}, \; \partial_\vartheta)$ of Grassmannian variables with the mappings: 
$s_d \longleftrightarrow \partial_\vartheta$ and  $s_{ad} \longleftrightarrow \partial_{\bar\vartheta}$ [cf. Secs. 4, 5].\\

\section 
{\bf Conclusions}

\vskip 0.5cm

In our present analysis, for the first time,  we have derived the off-shell nilpotent {\it quantum} (anti-)BRST along with 
(anti-)co-BRST symmetry transformations within the scope of ACSA. We have also discussed
the nilpotency along with  absolute anti-commutativity  properties of the  corresponding (anti-)BRST 
along with  (anti-)co-BRST  conserved  charges of the {\it ordinary} (0 + 1)-dimensional gauge
 invariant Christ--Lee model within the ambit  of (anit-)chiral supervariable  approach (ACSA) to BRST formalism.

The {\it novel} remarks of our present endeavor are the derivation  of the  off-shell nilpotent (anti-)BRST along with  (anti-)co-BRST 
symmetry transformations (cf. Sec. 4) and  the proof of nilpotency and  
the anti-commutativity properties  of the (anti-)BRST and (anti-)co-BRST charges in spite of  the fact that 
we have taken into account only the (anti-)chiral super expansions of the supervariables (cf. Sec. 5).
The nilpotency and anti-commutativity properties of the above conserved charges and derivation of the  
corresponding (anti-)BRST and (anti-)co-BRST symmetry transformations are obvious when the {\it full} super
expansions of the supervariables (i.e. BT-supervariable formalism  [29-32]) is taken into account. However, for the present study, 
 we have shown these properties 
with the help  of  {\it only} (anti-)chiral super expansions of the  (anti-)chiral supervariables.

It is worthwhile to mention that the nilpotency of the BRST as well as  anti-BRST  conserved 
charges is connected with the nilpotency ($\partial_{\bar\vartheta}^2 = \partial_\vartheta ^2 = 0)$ of the translational generators 
$\partial_{\bar\vartheta}$ and $\partial_\vartheta$, respectively. On the other hand, nilpotency of the co-BRST and anti-co-BRST
charges is connected with  nilpotency ($ \partial_\vartheta ^2  = \partial_{\bar\vartheta}^2 = 0$)    
of the translational generators $\partial_\vartheta$ and  
$\partial_{\bar\vartheta}$,   respectively. However, we have shown (cf. Sec. 5) that the absolute anti-commutativity
of the BRST charge   with anti-BRST charge  is connected with the nilpotency ($\partial_{\vartheta}^2 = 0$)
 of the translational generator ($\partial_\vartheta$) and absolute anti-commutativity of anti-BRST charge with BRST charge 
is connected with the nilpotency  ($\partial_{\bar\vartheta}^2 = 0$)  of the translational  generator ($\partial_{\bar\vartheta}$).
On the contrary, the absolute anti-commutativity of the co-BRST charge  with anti-co-BRST charge is connected with the nilpotency 
 of the translational generator ($\partial_{\bar\vartheta}$) and the absolute anti-commutativity of the anti-co-BRST 
charge with co-BRST charge is deeply related  with the nilpotency  of the translational generator ($\partial_{\vartheta}$).
 These statements  are completely novel for the present model.  
We have also captured the (anti-)BRST along with (anti-)co-BRST invariances of the 
Lagrangian within the scope  of ACSA. In fact, the action corresponding to the 
(anti-)chiral super Lagrangian is  independent of Grassmannian variables 
$(\vartheta, \bar\vartheta)$ which is completely  {\it novel}  for the present CL model (cf. Sec. 6).

The above issues, within the scope  of ACSA to BRST approach,  would be discussed in our future investigations for the
various gauge-invariant  models/theories like ABJM theory [33-35], supersymmetric Chern-Simons theory [36],
Jackiw-Pi model, Freedman-Townsend model, Abelian gauge theory with higher derivative matter fields.
In fact, our standard techniques of ACSA to BRST formalism are applicable wherever gauge invariance is present in the theory. 
Furthermore, there is a interesting and important work [37] that would be discussed in the future for different 
prospects of the theoretical and physical point of view in the domain of theoretical high energy physics. \\

\vskip 1.6cm

\noindent
{\bf\large Data Availability}\\

\noindent
No data were used to support this study. \\

\vskip 0.6 cm

\noindent
{\bf\large Conflicts of Interest}\\

\noindent
The authors of this study declare that there is no conflicts of interest.\\

\vskip 0.9 cm

\noindent
{\bf Acknowledgments:}
B. Chauhan and S. Kumar are  thankful to the DST-INSPIRE and BHU  fellowships for financial support, 
respectively. The authors  also thank  Dr. R. Kumar for a careful reading of the manuscript and for
important as well as significant suggestions.\\

\vskip 0.8 cm

\noindent
{\bf\large arXiv identifier}\\

\vskip 0.2 cm

\noindent 
arXiv: 2102.03845 [hep-th]\\

\vskip 0.5cm

\noindent

\end{document}